\documentclass{article}

\usepackage{arxiv}

\usepackage[utf8]{inputenc} 
\usepackage[T1]{fontenc}    
\usepackage{hyperref}       
\usepackage{url}            
\usepackage{booktabs}       
\usepackage{amsfonts}       
\usepackage{nicefrac}       
\usepackage{microtype}      
\usepackage{lipsum}

\usepackage{times}  
\usepackage{helvet} 
\usepackage{courier}  
\usepackage{graphicx}  
\usepackage{amsmath}
\usepackage{float}
\usepackage{subcaption}

\title{Agent Probing Interaction Policies}

\author{ \Large \textbf{Siddharth Ghiya\textsuperscript{\rm 1}, Oluwafemi Azeez\textsuperscript{\rm 1}, Brendan Miller\textsuperscript{\rm 1}}\\ 
\textsuperscript{\rm 1} 
Carnegie Mellon University\\
Pittsburgh, PA, USA\\ 
}

\begin{document}
\maketitle

\begin{abstract}
Reinforcement learning in a multi agent system is difficult because these systems are inherently non-stationary in nature. In such a case, identifying the type of the opposing agent is crucial and can help us address this non-stationary environment. We have investigated if we can employ some probing policies which help us better identify the type of the other agent in the environment. We've made a simplifying assumption that the other agent has a stationary policy from a fixed set that our probing policy is trying to classify. Our work extends Environmental Probing Interaction Policy framework to handle multi agent environments.
\end{abstract}

\section{Introduction}
Real world environments often have multiple agents. In such a case, considering the actions of the other actors in the environment becomes crucial. In many of the scenarios, the identity of other agents may not be known. Therefore, identifying the type of other agent is useful so that we can make a more informed decision taking into account the nature of behaviour of other agents in the environment.

Tremendous progress has been made in the field of multi-agent reinforcement learning. However most of these works have considered scenarios which are either only collaborative or scenarios where the type of the opposing agent is already known. In real world scenarios, agents often have to work in environments where there might be other unknown agents present. In such scenarios, we can perform actions and observe how the other agent in the environment reacts to our actions to determine the type of the other agent.

In this work, we present a framework which combines a Reinforcement Learning agent and an LSTM classifier to identify the type of the opposing agent present in the environment. We learn a "probing" policy which our agent performs, and observe the reactions of the other agent in the environment.

\section{Related Work}
\subsection{Multi Agent reinforcement learning}
MARL is one of the more widely studied topics in the field of reinforcement learning. Independent Q-Learning \cite{DBLP:journals/corr/TampuuMKKKAAV15} represents some of the earliest works in this field. In Independent Q-Learning, agents in the environment are trained with the assumption that the other agent is part of the environment. Naturally, this fails if we increase the number of agents in the environment due to non-stationarity in the environment. \\
More recently, there has been some work in MARL which can be classified under the "centralised training and decentralised execution paradigm" \cite{DBLP:journals/corr/LoweWTHAM17} \cite{DBLP:journals/corr/FoersterFANW17} \cite{DBLP:journals/corr/SunehagLGCZJLSL17}\cite{DBLP:journals/corr/abs-1803-11485}. In most of these works, a centralised critic is maintained during the training process which accounts for the non-stationarity in the environment. In all of these works, no communication is assumed between the agents. There has been some work in which agents learn to communicate with each other \cite{sukhbaatar2016learning} \cite{hoshen2017vain} \cite{foerster2016learning} \cite{mordatch2017emergence} which helps counter the non-stationarity of the environment. Some of the works don't explicitly make assumptions about the type of other agents in the environment. In this context, there has been work where an agent actively tries to modify the learning behaviour of the opposing agent \cite{foerster2017learning} or employs recursive reasoning to reason about the behaviour of other agents in the environment \cite{wen2018probabilistic}. \\
Our work falls more under the paradigm of agent modelling \cite{Albrecht_2018}. \cite{grover2018learning} \cite{alshedivat2017continuous} try to reason about the type of the other agent in the environment by observing their behaviour while carrying out their own policy. As they collect more data about the opposing agent in the environment, they condition their policy on the belief of the other agent and are able to achieve better generalization against different kinds of agents. To the best of our knowledge, in none of the available literature, attempts have been made to learn a policy which helps to identify the type of other agent present in the environment. To this end, we want to propose a separate policy, which when executed helps identify the type of the other agent in the environment. We speculate that, having information about the type of the other agent in the environment might help us in conditioning our policy and generalise better against different types of agents.

\subsection{Non stationary environments}
Multi-Agent systems are inherently non-stationary in nature. A lot of literature is available for dealing with non-stationary environments in single agent reinforcement learning. Many of these approaches use meta learning \cite{finn2017modelagnostic} \cite{wang2016learning} \cite{duan2016rl2} \cite{yu2018oneshot} to adapt to such non stationary environments. \cite{yu2017preparing} explicitly identify the properties of the environments which make it non-stationary and then use the obtained information to condition their policy. A more recent approach \cite{zhou2019environment} on dealing with non-stationary environments is learning a probing policy to identify the dynamics of the environment and then using this information to make an informed decision. We intend to use similar probing policies to help us better identify the type of other agent present in the environment.

\subsection{Classifier}
Our work does classification on trajectories, which are sequences of states. Recurrent neural networks are suitable for working with sequences. Long Short-Term Memory \cite{Hochreiter1997LongSM} is probably the most popular type of RNN and is used in our classifier. \\
Gated Recurrent Units \cite{Cho2014LearningPR} are an improved and more computationally efficient RNN. We may experiment with replacing our LSTM based classifier with a GRU in future works.

\section{Method}
Our approach involves iteratively training a probing RL agent and a classifier in alternating phases, as diagrammed in Fig. \ref{fig:overall_architecture}. First, the RL agent (see Fig. \ref{fig:api_agent_network_architecture}) executes a series of steps over several episodes according to a probing policy, while receiving reward based on a fixed classifier. The reward given to the RL agent during training is the negative cross entropy loss of the classifier. Then, after the probing policy is updated, a new classifier is trained.

\begin{figure*}
    \centering
    \includegraphics[width=\linewidth]{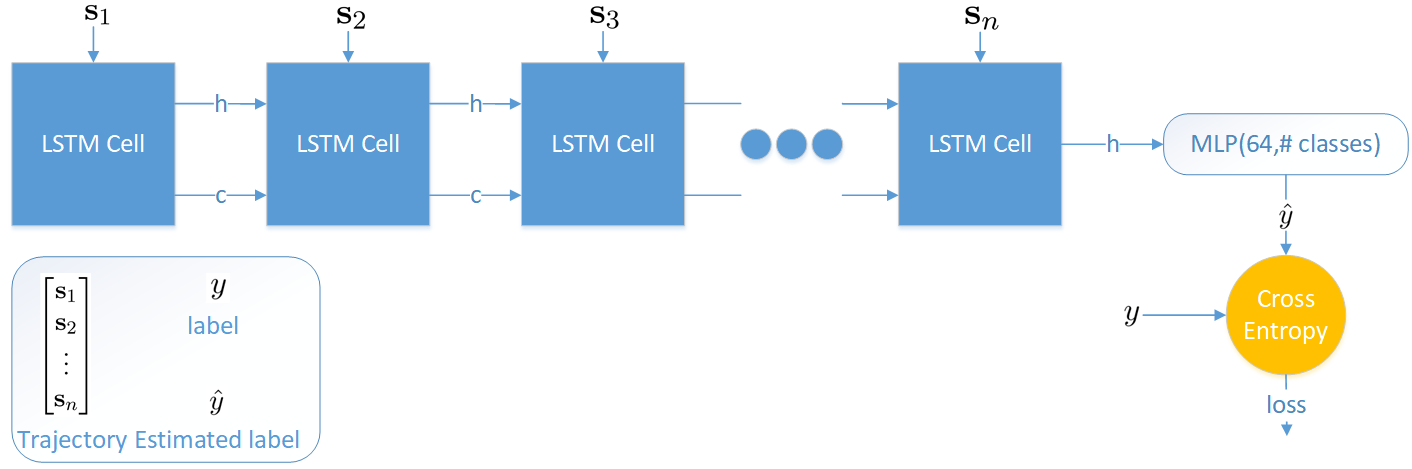}
    \caption{LSTM Based Classifier architecture}
    \label{fig:classifier}
\end{figure*}

During the classifier (see Fig. \ref{fig:classifier}) training phase, a large batch of trajectories are generated from the probing policy, along with the ground truth of the opposing agent type. The LSTM based classifier then learns to categorize the trajectories according to agent type. After the classifier is trained, it is frozen, and the RL agent training phase starts up again. This whole process repeats for several iterations.

The probing RL agent learns to take steps that can help the classifier detect the opposing agent correctly. Typically, this involves moving into the opposing agents field of view in order to cause it to take some action that can be used for classification. The opposing agent itself is controlled by a policy randomly selected at the beginning the episode.

Our current framework has no concept of an early done state. We always train with fixed length episodes.

\subsection{Environment}
\begin{figure}[H]
    \centering
    \includegraphics[width = 8cm]{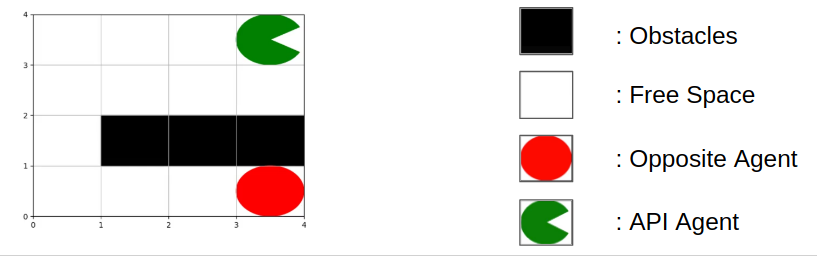}
    \caption{Explanation of the Environment}
    \label{fig:world}
\end{figure}
The environment used is a modification of openAI gym \cite{brockman2016openai} frozen lake environment. The environment was originally designed for a single agent navigating a grid world. The game ends whenever the agent either reaches a goal or falls into a hole. For the purpose of our experiment we modified the environment to use the holes as obstacles. Also the game has no concept of goal, it's only goal is to probe its opponent to help classify it properly. We also made the environment multi agent, i.e the agent and opponent agent take actions simultaneously. The state representation was designed to be easily to be used as input to the RL agent and the classifier. We used a matrix of size 4x4 with a value of 1 wherever there's an obstacle, 2 for where the agent is located and 3 for where the opponent is located. Rendering is done on the command line. 

\subsection{Opposing Agent}
We have two separate sets of experiments. In first set of experiments, we only have two kind of opposing agents in the environments, but we have a more challenging randomized environment. We also restrict the field of view of opposing agents. In the second set of experiments, we have many types of opposing agents, but the environment is simplified. Different settings of the opposing agents have been explained in more details below:

\subsubsection{Agents for Difficult Environments}

In this set of experiments, we only have two kinds of opposing agents in the environment. These opposing agents either go towards our API agent (predator) or go in the opposite direction (prey). Also, the opposing agent can only sense the presence of the API agent in the environment if the API agent is within its line of sight i.e. in the same row or in the same column with no intervening obstacles.

The starting positions of the agents are randomized but are always in positions where the API agent has to actively try to carry out a policy to try to get the opposing agent to react. A good example of such initialisation can be found in Fig. \ref{fig:world}. In this example, our API agent will have to navigate into the line of sight of the the opposing agent in order to get it to react.

\subsubsection{Agents for Many Agent Environments}
In this set of experiments, the opposing agent's policy is dependent on the main agent's actions. An agent can take four actions in the frozen lake gridworld, which are LEFT, RIGHT, UP and DOWN. The opposing agent can be deterministic or stochastic. when it's behaviour is deterministic, it can be 
\begin{enumerate}
    \item diagonal: This opposing agent would always take 90 degree anti-clockwise direction of the agent's direction. So if the agent goes LEFT, this agent would go DOWN.
    \item opposite: This opposing agent would always go the opposite direction of the original agent. So if the agent goes LEFT, then it would go RIGHT.
    \item opposite-diagonal: This opposing agent would always take 90 degree clockwise direction of the agent's direction. So if the agent goes LEFT, this agent would go UP.
    \item follower: This opposing agent always follow the original agent. So if the agent goes LEFT, this agent would always go LEFT.
\end{enumerate}

\begin{figure}[H]
    \centering
    \includegraphics[width=6 cm]{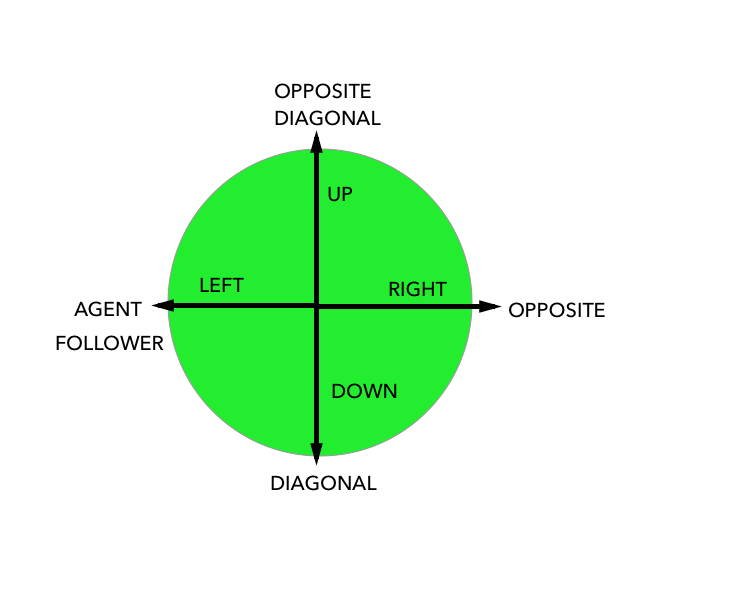}
    \caption{Opposing agents}
    \label{fig:opp}
\end{figure}

The opposing agents can also be stochastic. This means an action can be categorically sampled from a distribution with more weights on each of the options listed above. i.e, stochastic diagonal would give more weights to the diagonal direction in the probability distribution sampled from while an opposing agent would have more weights on the opposite direction of the agent. We then have a final opposing agent which just behaves randomly. This framework of opposing agents could be generalized to a scenario where we can always create $2n + 1$ opposing agent given n actions the original agent can take. The conjecture is that stochastic opposing agents would prove very challenging to the agent, it would be easier to detect deterministic agents than stochastic ones.

\subsection{Classifier}
The classifier takes in a trajectory specified as a world map $\mathbf{W}$ and a sequence of states $\mathbf{s}_1$ and $\mathbf{s}_2$ for agents 1 and 2.
The world $\mathbf{W}$ is an m by m occupancy grid specified as a binary matrix.

\begin{align}
    \mathbf{W} &= \begin{bmatrix}
    0 & 1 & \cdots & 0 \\
    \vdots &  \vdots & \vdots & \vdots \\
    1 & 0 & \cdots & 0 \\
    \end{bmatrix}
\end{align}

The states are sequences of row $r_i$ and column $c_i$ coordinates:

\begin{align}
    \mathbf{S}_1 = \begin{bmatrix}
    r_{11} & c_{11} \\
    r_{12} & c_{12} \\
    \vdots & \vdots \\
    r_{1n} & c_{1n}
    \end{bmatrix} \\
    \mathbf{S}_2 = \begin{bmatrix}
    r_{21} & c_{21} \\
    r_{22} & c_{22} \\
    \vdots & \vdots \\
    r_{2n} & c_{2n}
    \end{bmatrix}
\end{align}

When feeding into the LSTM based classifier the complete state for time $i$ is flattened and fed into the classifier, along with the world map of that state. Currently, the world map is fed in for each time step in the sequence, which is somewhat wasteful.

The complete sequence is described in $\mathbf{U}$ below:

\begin{equation}
     \mathbf{U} = \begin{bmatrix}
    r{11} & c{11} & c{21} & {c21} & w_{11} & w_{12} & \cdots & w_{mm} \\
    \vdots & \vdots & \vdots & \vdots & \vdots & \vdots &  & \vdots \\
    r{1n} & c{1n} & c{2n} & {c2n} & w_{11} & w_{12} & \cdots & w_{mm} \\
    \end{bmatrix}
\end{equation}

\subsection{API Agent}
\begin{figure*}
    \centering
    \includegraphics[width=\linewidth]{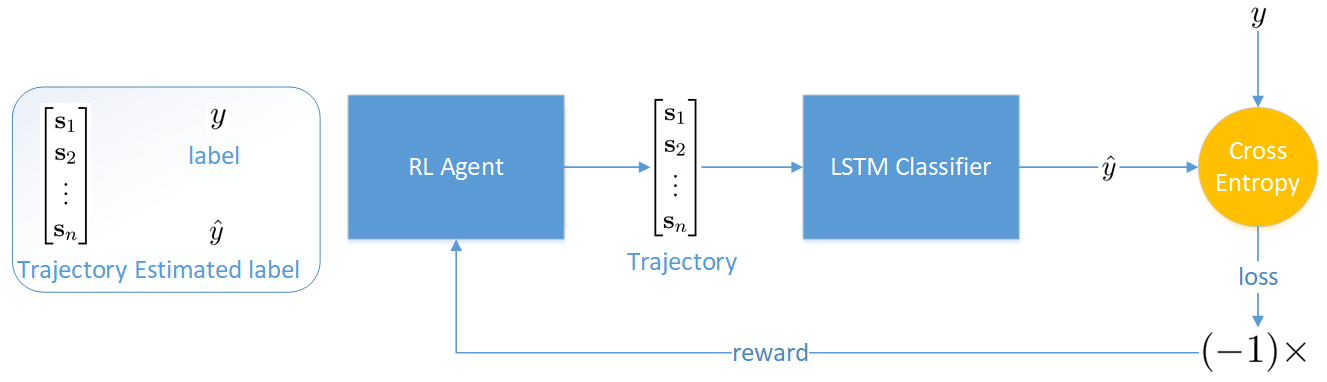}
    \caption{Diagram of the overall architecture}
    \label{fig:overall_architecture}
\end{figure*}

We use Proximal Policy Optimization Algorithm \cite{schulman2017proximal} for training the agent to follow a probing policy. The reward for the agent is the negative cross entropy loss of the classifier as shown in Fig. \ref{fig:overall_architecture}. Currently, we are using a sparse reward formulation in which the agent is rewarded only at the last time step.

\begin{figure}
    \centering
    \includegraphics[width=\linewidth]{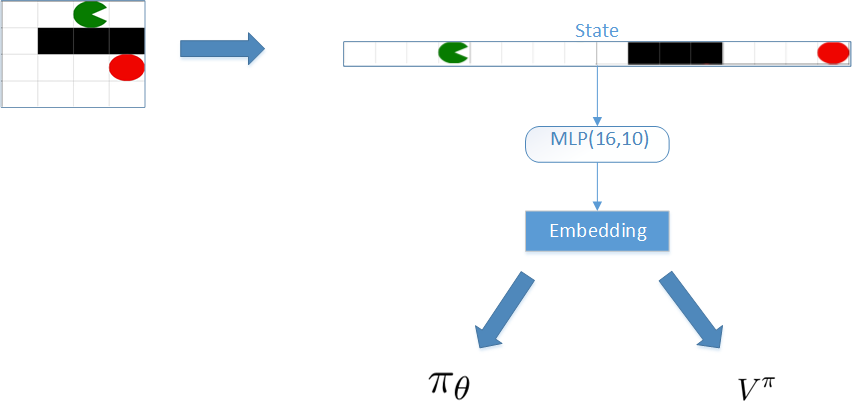}
    \caption{API Agent network architecture}
    \label{fig:api_agent_network_architecture}
\end{figure}

We have a simple 2 layer fully connected layer which takes in the current state of the world. We have two separate heads for policy and value. Please refer to Fig. \ref{fig:api_agent_network_architecture} for the network architecture of the probing agent.

\section{Results}
\subsection{Agents for Difficult Environments}
We have a trajectory classifier implemented using an LSTM that runs on the sequence of states. It runs on a fixed world pictured in Fig. \ref{fig:world}. In this experiment, a 4x4 world is used with some blocks in the center which must be navigated around.

Every episode starts with randomly placed agents. The agents then move according to their policy. In this experiment, the probing agent moves according to it's policy. It's then given a reward at the last time step of the trajectory. Currently we have a trajectory of maximum 20 time steps. The reward given to the API agent at the end of it's trajectory is negative of the loss of the classifier for that trajectory. Basically, the classifier looks at the sequence of states of the trajectory and outputs a label of the opposing agent. 

\begin{figure}
    \centering
\begin{minipage}{.4\textwidth}
    \centering
    \includegraphics[width=8cm]{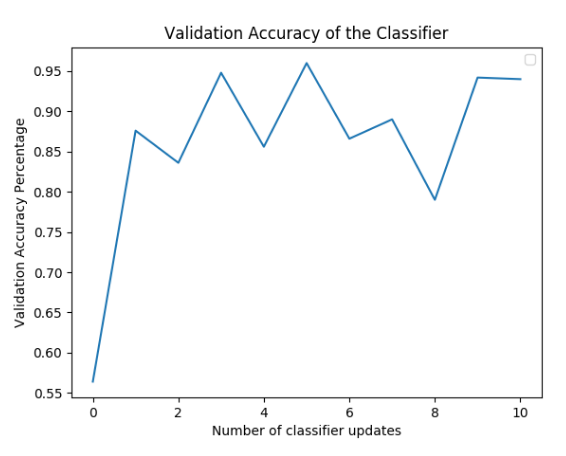}
    \caption{Accuracy of the classifier tested every 200 episodes}
    \label{fig:classifier_accuracy}
\end{minipage} \hfill%
\begin{minipage}{.4\textwidth}
    \centering
    \includegraphics[width=8cm]{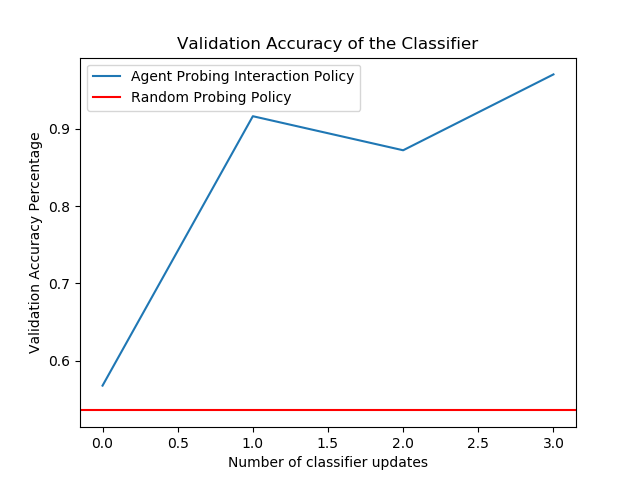}
    \caption{Comparison with random probing agent}
    \label{fig:random_probing}
\end{minipage}
\end{figure}

We randomly generate 1000 episodes of 20 time steps each. The API agent is trained in every episode. The classifier is trained only every 200 episodes. Also, whenever the classifier is trained, we generate some validation trajectories and calculate the accuracy for which the classifier is able to correctly identify the opposing agent in the environment. The environment which we used in these set of experiments is similar to world pictured in \ref{fig:world} with randomized obstacle positions and randomized start positions of the two agents. As we see in the results, the classifier is able to correctly classify trajectories and recognize the opposing agent(Fig. \ref{fig:classifier_accuracy}). 

We have a video of our experiments at : \url{https://www.youtube.com/watch?v=6Y5fpdrQifA}. In the video, it can be seen that the probing agent sometimes wanders off the optimal path to the probing position.

We think that we observe such a performance because currently we are providing reward to the API agent(negative loss of the classifier) at only the last time-step(which is fixed at 20). This can be improved by giving a reward to the agent at every time-step or by penalising its movement.

We also did a comparison with a random probing agent. In case of random probing agent, the probing agent is taking random actions at every time step. We then pass this trajectory through the classifier. As expected, in such a case, the classifier is not able to properly classify the opposing agent in the environment. Since there is no policy which guides the probing agent to a spot in the environment such that the opposing agent is able to see it and react to it's presence, the classifier ends up giving predicting random labels for the opposing agent(Fig. \ref{fig:random_probing}).

\subsection{Experiments with Many Agent Environment}
\subsubsection{One State Trajectory}
We randomly generated 1000 episodes of 128 time steps each. Then we train our network using the Adam optimizer with a learning rate of 0.001 for 20 epochs.

Each step was treated as a one trajectory move and the 128 steps as 128 batch. Average reward of the RL agent and Loss of the classifier of the 128 batch was then plotted against 1000 timestep. An experiment was setup for only deterministic opposing agents, deterministic and stochastic opposing agents and an agent that runs random policy was used as a baseline.

\begin{figure}
    \centering
\begin{minipage}{.4\textwidth}
    \centering
    \includegraphics[width=8cm]{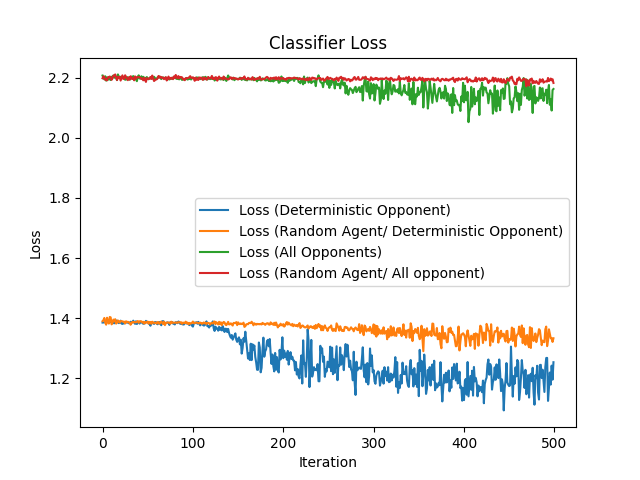}
    \caption{One state Trajectory Classifier Loss}
    Classifier Loss against number of iterations.
    \label{fig:loss}
\end{minipage}
\hfill%
\begin{minipage}{.4\textwidth}
    \centering
    \includegraphics[width=8cm]{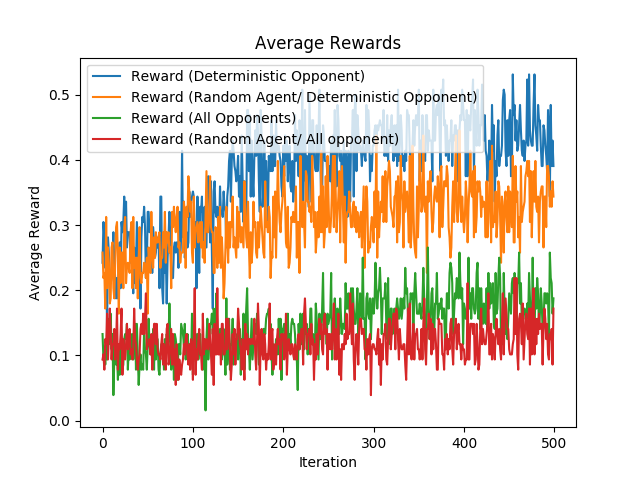}
    \caption{One state Trajectory Reward}
    Reward against number of iterations.
    \label{fig:reward}
\end{minipage}
\end{figure}

The classifier loss result is shown in Fig. \ref{fig:loss} and agent reward is shown in Fig. \ref{fig:reward}. It is obvious that classifier loss  reduces while the agent reward increases and is better than random in both case. The agent could easily classify the opposing agent whenever it's a simple as being deterministic, however when some of the opposing agent start behaving stochastically, it begins to get very difficult for the agent to classify. This generally exposes a rare kind of supervised learning difficulty where ground truth label data is stochastic. The loss of combination of all opponents in Fig. \ref{fig:loss} is higher than for deterministic opponents because the classification loss is for 9 opposing agent and would be as high as $\ln(9)$ initially compared to 4 opposing agent which starts at about $\ln(4)$. 
\subsubsection{128 State Trajectories}

To verify if adding extra information would provide extra information, we added another experiment in which we randomly generated 1000 episodes of 128 time steps each. Then we train our network using the Adam optimizer with a learning rate of 0.001 for 20 epochs.

Each step was treated as a 128 trajectory move and a batch size of 10. Average reward of the RL agent and Loss of the classifier of the 10 batch was then plotted against 1000 timestep. An experiment was setup for only deterministic opposing agents, deterministic and stochastic opposing agents and an agent that runs random policy was used as a baseline.

\begin{figure}
    \centering
\begin{minipage}{.4\textwidth}
    \centering
    \includegraphics[width=8cm]{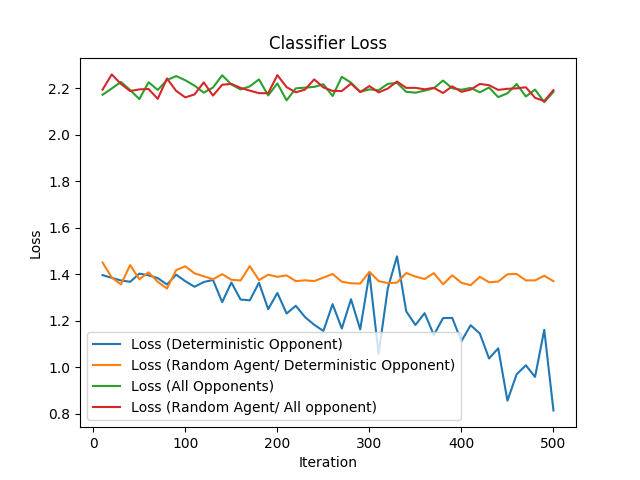}
    \caption{128 state Trajectory Classifier Loss}
    Classifier Loss against number of iterations.
    \label{fig:multiloss}
\end{minipage}
\hfill%
\begin{minipage}{.4\textwidth}
    \centering
    \includegraphics[width=8cm]{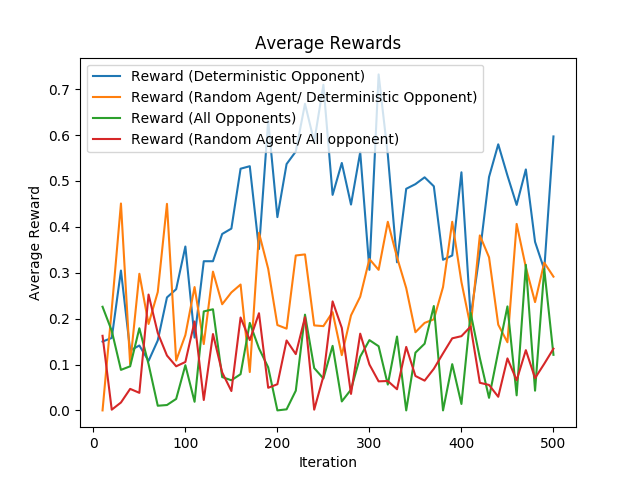}
    \caption{128 state Trajectory Reward}
    Reward against number of iterations.
    \label{fig:multireward}
\end{minipage}
\end{figure}

 The classifier loss result is shown in Fig. \ref{fig:multiloss} and agent reward is shown in Fig. \ref{fig:multireward}. The expectation here is that giving more state information in terms of a full trajectory to the LSTM should significantly improve the classification loss and agent reward. The behaviour is similar to the one state scenario. An observation is that the extra information seem to hurt the agent's performance whenever stochastic agents are considered. This can be interpreted as extra information introducing confusion to the classifier. 

\subsubsection{One State vs 128 State Trajectories}

\begin{figure}
    \centering
\begin{minipage}{.4\textwidth}
    \centering
    \includegraphics[width=8cm]{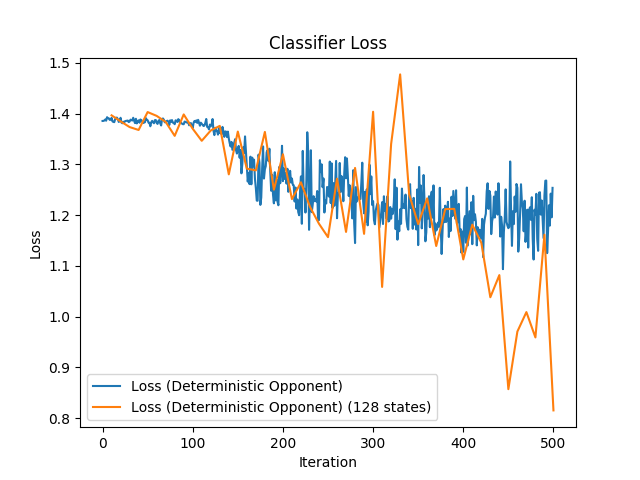}
    \caption{128 state Trajectory Classifier Loss}
    Classifier Loss against number of iterations.
    \label{fig:compareloss}
\end{minipage}
\hfill%
\begin{minipage}{.4\textwidth}
    \centering
    \includegraphics[width=8cm]{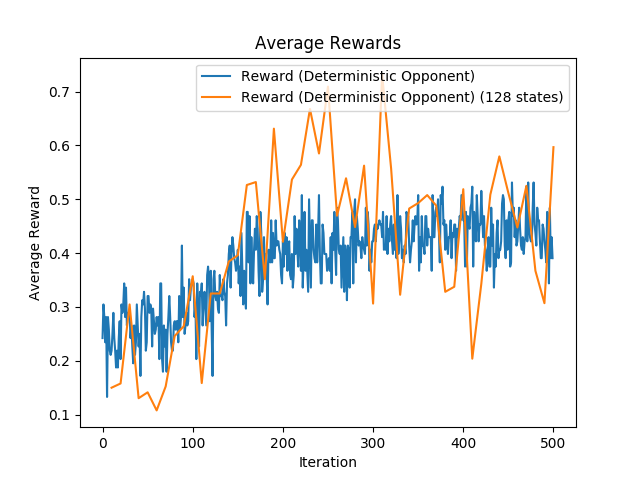}
    \caption{128 state Trajectory Reward}
    Reward against number of iterations.
    \label{fig:comparereward}
\end{minipage}
\end{figure}

The one state and 128 state trajectory approaches were compared and the result presented in fig \ref{fig:compareloss} and fig \ref{fig:comparereward}. The observation is that the 128 state trajectory performed better than one state trajectory when only deterministic opposing agents are considered, even though variance was generally more. This implies that providing more information to the agent does not significantly improve the performance which may be because of the simplicity of the environment. There was no improvement when stochastic opposing agents were added, it is plausible to notice that extra information about a stochastic opposing agent might prove to be much more confusing. 

\section{Ethics}

The key ethical risk of the use of agent probing policies in a real-world environment is one of instrumentation causing harm. The probing policy is trained to determine the opposing agents type by any means. The probing policy does not take into account the welfare of either agent.

To make this more concrete, consider an autonomous driving scenario. In this scenario, we want to know if other drivers on the road are distracted. A distracted driver could be defined as one who will not react swiftly to a dangerous driving situation such as an obstacle in its path. In this scenario, the easiest way to determine whether a driver will react to a dangerous situation is to create a dangerous situation and observe how the driver reacts. Likely, the probing agent will simply drive directly at the opposing agent and see if the opposing agent swerves to avoid the probing agent.

We can see this behavior in our experiments where we have restricted the field of vision of the opposing agent. The probing policy simply moves directly into the field of view of the opposing agent to see how it will react. In our experiments, we have defined predator and prey opposing agents, and the probing policy does not consider the danger of encountering a predator agent.

To solve this problem, the process for training the probing policy could be modified so that it receives negative reward if it takes a dangerous action. However, a practical solution to this problem is saved for future work.

A secondary ethical issue of the probing policy relates to regularity and accountability. Consider the autonomous driving scenario outlined above. For the reasons discussed, it may not be possible to safely determine whether another driver is distracted or not. For this reason, the regularity of the classifiers ability to determine the opposing agent type is low. This lack of regularity means that manufacturer of the autonomous driving system can’t accept responsibility for the safety of the system in the presence of distracted drivers. Thus, the manufacturer lacks accountability.

\section{Conclusions and Future Work}
\subsection{Improvements with Current Experiments}
We want to do the following improvements with our current experiment setup:
\begin{enumerate}
\item  In the work mentioned in this paper, we were able to show that our agent was indeed able to learn a probing policy. It displayed behaviour of trying to go to a position in the environment to observe the reaction of the opposing agent. However, we also noticed that the agent sometimes seemed to wander off the optimal trajectory and take random steps. This is because we are only giving it a sparse reward at the end of the trajectory. In our current experiments, the trajectory is of fixed length and therefore the agent does not learn to go to the probing position as soon as possible. A simple fix to this problem can be providing the negative of the loss of the classifier at every time step to the agent. This might help the agent learn to go to the probing position in the environment optimally.
\item  In the current experiments, we have a fixed number of time steps to execute our probing policy. If we want to use the type of the opposing agent in the environment to actually help us in our task policy, we need to have a done condition for our probing policy. If the classifier is able to classify the opposing agent as a certain type of agent with certainty above a threshold percentage, we can stop executing our probing policy and start executing our task policy. We can modify our current experiments to add such an early done condition.
\end{enumerate}

\subsection{Further Future Directions}
In current experiments, we are giving the label of the opposing agent in the environment as a supervised target for calculating loss and training the classifier and the API policies. In real world scenarios, we would ideally want our agent to form a belief about the opposing agent in the environment and use this belief to condition its task policy. We would want our agent to learn such a belief about the opposing agent without supervision. A promising future direction can be coming up with an architecture for such end to end training of API policies. 


\bibliographystyle{unsrt}
\bibliography{output}

\end{document}